\documentclass[12pt]{iopart}
\usepackage{graphicx}
\begin{document}

\hfill  DAMTP-2015-33

\title{Twistor form of massive 6D superparticle}

\author{Alasdair J. Routh and Paul K. Townsend}
\address{Department of Applied Mathematics and Theoretical Physics,\\ Centre for Mathematical Sciences, University of Cambridge,\\
Wilberforce Road, Cambridge, CB3 0WA, U.K.}
\ead{A.J.Routh@damtp.cam.ac.uk, P.K.Townsend@damtp.cam.ac.uk}
\vspace{10pt}
\begin{abstract}
The massive six-dimensional (6D)  superparticle with manifest (n,0) supersymmetry is shown to have a 
supertwistor formulation in which  its ``hidden''  (0,n) supersymmetry is also manifest. The mass-shell constraint is replaced by 
Spin(5)  spin-shell constraints which imply that the quantum superparticle has zero superspin; for n=1 it  propagates the 6D Proca supermultiplet. 
\end{abstract}

%
%
%
%
%

\font\mybb=msbm10 at 12pt
\def\bb#1{\hbox{\mybb#1}}
\def\bZ {\bb{Z}}
\def\bR {\bb{R}}
\def\bE {\bb{E}}
\def\bT {\bb{T}}
\def\bM {\bb{M}}
\def\bL {\bb{L}}
\def\bG {\bb{G}}
\def\bH {\bb{H}}
\def\bC {\bb{C}}
\def\bA {\bb{A}}
\def\bK {\bb{K}}
\def\bO {\bb{O}}
\def\bX {\bb{X}}
\def\bP {\bb{P}}
\def\bJ {\bb{J}}
\def\bU {\bb{U}}
\def\bJ {\bb{J}}
\def\bV{\bb{V}}
\def\bW {\bb{W}}
\def\bI {\bb{I}}

\def\bfo{\mbox{\boldmath $\omega$}}
\def\bfO{\mbox{\boldmath $\Omega$}}
\def\bfn{\mbox{\boldmath $\nabla$}}
\def\bfs{\mbox{\boldmath $\sigma$}}
\def\bft{\mbox{\boldmath $\tau$}}
\def\bfS{\mbox{\boldmath $\Sigma$}}
\def\bfpsi{\mbox{\boldmath $\psi$}}
\def\bfG{\mbox{\boldmath $\Gamma$}}
\def\bfth{\mbox{\boldmath $\theta$}}
\def\bfTh{\mbox{\boldmath $\Theta$}}
\def\bfPi{\mbox{\boldmath $\Pi$}}
\def\bfpi{\mbox{\boldmath $\pi$}}
\def\tr{{\rm tr}}


\newcommand{\sect}[1]{\setcounter{equation}{0}\section{#1}}
\renewcommand{\theequation}{\arabic{section}.\arabic{equation}}

\section{Introduction}
\setcounter{equation}{0}

Twistors are spinors of  (a cover of) the conformal group. They arise in formulations of conformally invariant theories that make  the conformal invariance manifest.  For spacetime dimensions $D=3,4,6$ (which we abbreviate to 3D etc.)  there is  a natural superconformal extension of the conformal group \cite{Nahm:1977tg} and hence a natural extension of twistors to supertwistors \cite{Ferber:1977qx}, which can be used to construct  manifestly superconformally invariant theories in these dimensions; recent field theory examples can be found in  \cite{Dennen:2009vk,Huang:2010rn}.  In the context of particle mechanics,  the superconformal invariance of the massless superparticle becomes manifest in a phase-space formulation in which the phase-space coordinates are the components of a supertwistor \cite{Shirafuji:1983zd, Bengtsson:1987ap,Bengtsson:1987si,Cederwall:1993xe}.

Surprisingly, twistor methods are not limited to massless particle mechanics, although a doubling of the twistor phase space is needed to allow for a non-zero mass  \cite{Hughston:1981zc}. 
One way to understand how it is that twistors can be relevant to massive particles is to consider a massive particle as a massless particle in a higher dimension. For example, by starting with the supertwistor form of the massless 6D superparticle action, a  double-supertwistor form of the action for a particular  4D massive superparticle  is found upon  imposing appropriate momentum constraints \cite{deAzcarraga:2008ik}.  A review of this idea, with extensions and other applications of it,  can be found in  \cite{Mezincescu:2013nta}. 

There is no analogous way to obtain a double-supertwistor  formulation of the massive 6D superparticle. Although the standard  massive 6D superparticle action can be found by imposing momentum constraints on the massless 10D superparticle,  there is no adequate supertwistor formulation of the latter that could be used to find the supertwistor formulation of the former; see e.g. \cite{Howe:1991hk,Berkovits:1990yc} for a discussion of the difficulties.  Nevertheless, a {\it direct} construction of a double-supertwistor formulation of the massive 6D superparticle is possible,  as we show here.  This construction could provide further insight into the massless 10D case, which is of  relevance to superstring theory \cite{Brink:1981nb}.

Apart from this possible link to superstrings, one may ask what advantages twistors have when there is no conformal invariance to be made manifest. One answer to this question emerged 
from the results of \cite{Mezincescu:2013nta} for the simplest  ${\cal N}=1$ massive 4D superparticle, which has a second ``hidden'' supersymmetry  implying an equivalence  to the 
${\cal N}=2$ massive ``BPS superparticle'' (which is directly related to the massless 6D superparticle)  \cite{Mezincescu:2014zba}. This equivalence becomes manifest when the twistor formulations of the two actions are compared: they  are identical!  

It was further shown in \cite{Mezincescu:2014zba} that this equivalence  is a general feature of   massive superparticle actions (in a Minkowski vacuum background) in any spacetime dimension
because non-BPS massive superparticle actions are just versions of a  BPS massive superparticle action for which the latter's fermionic gauge invariance (``kappa-symmetry'') has been (partially or fully) 
gauge-fixed. The gauge fixing preserves manifest Lorentz invariance but obscures some of the supersymmetries. For example, one can gauge-fix the 6D massive BPS superparticle action with manifest 
$(n,n)$ supersymmetry to arrive at a much simpler 6D massive superparticle action with no fermionic gauge invariance; this action still has $(n,n)$ supersymmetry, of course, but only the $(n,0)$ supersymmetry 
is now  manifest.  This result  greatly simplifies our present task because it allows us to focus {\it without loss of generality}  on  massive superparticle actions  without fermionic gauge invariances. 

For example, the simplest such action for a superparticle of mass $m$ is, in phase-space form, 
\begin{equation}\label{saction}
S= \int dt \left\{ \left[\dot X^m + i \left(\bar\Theta\Gamma^m\dot\Theta - \dot{\bar\Theta} \Gamma^m\Theta\right)\right]P_m - \frac{1}{2}e\left( P^2 +m^2\right)\right\}\, , 
\end{equation}
where $\Theta$ is a complex chiral anticommuting  6D spacetime spinor, and $e(t)$ is the Lagrange multiplier for the mass-shell constraint (we assume a Minkowski spacetime metric with ``mostly plus'' signature and coordinates $\{X^m ;m=0,1,\dots,5\}$). This action has manifest $(1,0)$ supersymmetry but also, for $m\ne0$,  a ``hidden''  $(0,1)$ supersymmetry \cite{Mezincescu:2014zba}. It is in canonical Hamiltonian form when $m\ne0$ because in this case  it defines an invertible closed (orthosymplectic) two-form  on the phase superspace with coordinates $(X,P,\Theta)$.

As we shall see, the  full $(1,1)$ supersymmetry of the action  (\ref{saction}) becomes manifest in its supertwistor form. This  involves a pair of 6D supertwistors, of the same chirality, on which there is a natural action of $USp(4)\cong {\rm Spin}(5)$, which emerges  as a gauge invariance of the supertwistor action, with corresponding  ``spin-shell' constraints.  Coincidentally, ${\rm Spin}(5)$ is also the 6D  rotation group,  which is Wigner's ``little group'' for massive particles in 6D.  In reality, this is no coincidence but it is not immediately obvious what the connection is between space rotations and the ``internal'' ${\rm Spin}(5)$ gauge group.  This issue was addressed for the massive 4D superparticle in \cite{Mezincescu:2013nta}, but here we present a simpler resolution of it, in the 6D context,  by consideration of the supersymmetric extension of the Pauli-Lubanski (PL) tensors. 

Pauli-Lubanski tensors are generalizations of the 4D Pauli-Lubanski  ``spin-vector''; they are translation invariant tensors constructed from the Poincar\'e Noether charges  $\{{\cal P},{\cal J}\}$. In 6D the PL tensors are
\begin{equation}\label{PLT}
\Sigma ^{mnp}= \varepsilon^{mnpqrs}{\cal J}_{qr}{\cal P}_s\, , \qquad \Xi^m=  \varepsilon^{mnpqrs}{\cal J}_{np}{\cal J}_{qr}{\cal P}_s\, . 
\end{equation}
In the context of classical  particle mechanics, the Poincar\'e Noether charges are tensors on phase space.  When these charges are expressed in terms of the usual phase space coordinates for a massve point particle, the  PL tensors are identically zero. This is no longer true in the  double-twistor formulation; instead, the PL tensors are zero {\it as a consequence of the spin-shell constraints}, so  these constraints imply that the particle has zero spin. Here we show that an analogous result holds for the 6D massive superparticle if the PL tensors are replaced by what we shall refer to 
as the super-PL tensors.  It turns out that all super-PL tensors are zero as a consequence of the superparticle spin-shell constraints, which implies that the quantum superparticle describes a 
massive supermultiplet of  zero superspin.  For $n=1$ this is the 6D Proca supermultiplet of maximum spin $1$, but to realize the full BPS-saturated $(1,1)$ supersymmetry it  must be ``centrally charged'', which implies a doubling of the states. 

Throughout this paper, we make extensive use of the $SU^*(4)$ notation for 6D Minkowski spinors  \cite{Kugo:1982bn,Howe:1983fr,Koller:1982cs}. We begin with a brief review of this notation as it applies to the particle and superparticle  in their  standard phase-space formulations. Then we present the twistor formulation of the bosonic 6D particle, followed by a generalization to the 6D massive superparticle with manifest $(n,0)$ supersymmetry, confirming its BPS-saturated $(n,n)$ supersymmetry. 

We conclude with a discussion of how the results obtained here fit into the general pattern of twistor formulations of particle mechanics models in $D=3,4,6$ spacetime dimensions,  and their relation to the division algebras $\bR,\bC,\bH$, and we comment on implications for the $D=10$ case in relation to the octonions $\bO$.

\section{6D preliminaries}
\setcounter{equation}{0}

In  $SU^*(4)$ notation, 6D vectors are anti-symmetric bi-spinors. In particular, the standard phase space coordinates for a point particle are 
$(\bX^{\alpha\beta},\bP_{\alpha\beta})$ ($\alpha,\beta=1,2,3,4$) and  the action for a particle of mass $m$ is
\begin{equation}\label{phase}
S= \int\! dt\left\{ \dot{\bX}^{\alpha\beta} \bP_{\alpha\beta} - \frac{1}{2} e\, \left(\bP^2+ m^2\right) \right\} \, ,  \qquad \left(\bP^2= \bP^{\alpha\beta}\bP_{\alpha\beta}\right). 
\end{equation}
As for all other Lorentz 6-vectors, we raise indices using the  alternating invariant tensor of $SU^*(4)$: 
\begin{equation}
\bP^{\alpha\beta} =  \frac{1}{2} \varepsilon^{\alpha\beta\gamma\delta} \bP_{\gamma\delta} \,  \qquad  \Rightarrow \quad
\bP^{\alpha\beta}\bP_{\alpha\gamma} = \frac{1}{4} \delta_\gamma^\beta\, \bP^2\, . 
\end{equation}
Similarly, 6-vector indices may be lowered using the inverse alternating invariant tensor of $SU^*(4)$, defined such that 
\begin{equation}
\frac{1}{2}\varepsilon^{\alpha\beta\epsilon\eta}\, \varepsilon_{\epsilon\eta\gamma\delta} = 2\, \delta^\alpha_{[\gamma}\, \delta^\beta_{\delta]}\, , 
\end{equation}
where the brackets indicate ``unit strength'' antisymmetrization over enclosed indices. 
We remark here, for future use, that if the spinor components of $\bP$ are interpreted as entries of a matrix $\bP$, then 
\begin{equation}\label{detp}
16 \det\bP = \left(\bP^2\right)^2\, . 
\end{equation}

The canonical Poisson bracket relations following from the action (\ref{phase}) are
\begin{equation}
\left\{\bX^{\alpha\beta},\bP_{\gamma\delta}\right\}_{PB} = \delta^\alpha_{[\gamma}\delta^\beta_{\delta]}\, . 
\end{equation}
The Poincar\'e Noether charges in spinor notation are
\begin{equation}\label{Noethertwist}
{\cal P}_{\alpha\beta} = \bP_{\alpha\beta}\, , \qquad 
{\cal J}_\alpha{}^\beta = 2\bP_{\alpha\gamma}\bX^{\beta\gamma} - \frac{1}{2}\delta_\alpha^\beta\left(\bP_{\gamma\delta}\bX^{\gamma\delta}\right)\, , 
\end{equation}
and their non-zero Poisson brackets are
\begin{eqnarray}
\left\{{\cal J}_\alpha{}^\beta, {\cal P}_{\gamma\delta}\right\}_{PB} &=& \delta^\beta_\gamma {\cal P}_{\alpha\delta} + 
\delta^\beta_\delta {\cal P}_{\gamma\alpha}- \frac{1}{2}\delta^\beta_\alpha {\cal P}_{\gamma\delta} \, , \nonumber \\
\left\{{\cal J}_\alpha{}^\beta, {\cal J}_\gamma{}^\delta\right\}_{PB} &=& \delta^\beta_\gamma {\cal J}_\alpha{}^\delta 
- \delta_\alpha^\delta {\cal J}_\gamma{}^\beta\, . 
\end{eqnarray}

\subsection{Pauli-Lubanski tensors}

As remarked in the introduction, there are two 6D analogs of the 4D Pauli-Lubanski spin vector. In $SU^*(4)$ spinor notation, the self-dual and anti-self-dual parts of the PL 3-form tensor are
\begin{equation}\label{sigpm}
\Sigma^{(+)}_{\alpha\beta} = {\cal J}_{(\alpha}{}^\gamma{\cal P}_{\beta)\gamma}\, , \qquad 
\Sigma_{(-)}^{\alpha\beta} = {\cal J}_\gamma^{(\alpha}{\cal P}^{\beta)\gamma}\, . 
\end{equation}
In the same spinor notation, the PL  vector $\Xi$  is\footnote{This corrects the expression given in \cite{Mezincescu:2013nta}.}
\begin{equation}\label{Xi}
\Xi_{\alpha\beta} \ = -2{\cal P}_{\delta[\alpha} {\cal J}_{\beta]}{}^\gamma {\cal J}_\gamma{}^\delta - \frac{1}{2} {\cal P}_{\alpha\beta}\left({\cal J}_\delta{}^\gamma {\cal J}_\gamma{}^\delta\right)\, . 
\end{equation}
To verify translation invariance of the PL tensors (i.e. that they have zero Poisson bracket with ${\cal P}$), one needs the  identities 
\begin{equation}
\bP_{[\alpha\beta}\bP_{\gamma]\delta} \equiv \frac{1}{12} \bP^2 \, \varepsilon_{\alpha\beta\gamma\delta} \, , \qquad
\varepsilon_{\eta\epsilon\gamma[\alpha}{\cal J}_{\beta]}{}^\gamma + \varepsilon_{\alpha\beta\gamma[\eta}{\cal J}_{\epsilon]}{}^\gamma \equiv 0\, .
\end{equation}

The Pauli-Lubanski tensors vanish identically when the Poincar\'e charges are expressed in terms of the phase-space variables $(\bX,\bP)$ but not 
when expressed in terms of the twistor phase-space variables to be introduced later. The PL tensors themselves satisfy the identities
\begin{equation}\label{PLids}
{\cal P}^{\alpha\gamma}\, \Sigma^{(+)}_{\gamma\beta} \equiv {\cal P}_{\beta\gamma}\, \Sigma_{(-)}^{\gamma\alpha}\, , \qquad {\cal P}^{\alpha\beta}\, \Xi_{\alpha\beta} \equiv 0\, ,  
\end{equation}
and the spinor relation expressing the fact that $\Xi$ is a contraction of ${\cal J}$ with $\Sigma$ is
\begin{equation}\label{XS}
\Xi_{\alpha\beta} =  \Sigma^{(+)}_{\delta[\alpha} {\cal J}_{\beta]}{}^\delta  - \frac{1}{2}\varepsilon_{\alpha\beta\gamma\delta} \Sigma_{(-)}^{\eta\gamma}{\cal J}_\eta{}^\delta\, . 
\end{equation}
The main reason for the importance of PL tensors, for massive particles,  is that the scalars constructed from them are proportional to Casimirs of the  Poincar\'e group. 
In 6D there are two such scalars\footnote{This count excludes the Casimir ${\cal P}^2$, which is not constructed from a PL tensor.}
\begin{equation}
\Sigma^2 = \Sigma^{(+)}_{\alpha\beta} \Sigma_{(-)}^{\alpha\beta}\, , \qquad \Xi^2 = \frac{1}{2} \varepsilon^{\alpha\beta\gamma\delta}\Xi_{\alpha\beta}\Xi_{\delta\gamma}\, . 
\end{equation}

\subsection{Massive superparticle}

The minimal 6D spinor is a complex ${\bf 4}$ of $SU^*(4)$, which can be traded for a $({\bf 2},{\bf 4})$ of $SU(2)\times SU^*(4)$ subject to a  ``symplectic reality condition''. More generally, 
a set of  $n$ such spinors of the same chirality naturally transform as the $({\bf 2n},{\bf 4})$ of  $USp(2n)\times SU^*(4)$, again subject to a  ``symplectic reality condition'' 
(see e.g. \cite{Kugo:1982bn}).  The $n$ minimal anticommuting  spinors needed for a 6D superparticle with $(n,0)$ supersymmetry thus combine to form a single  spinor $\Theta^\alpha_i $  ($i=1,\dots 2n$) which has $4n$ independent {\it complex} components.  Using this notation, the action for the massive 6D 
superparticle with manifest $(n,0)$ supersymmetry is 
\begin{equation}
S= \int\! dt\left\{ \left(\dot{\bX}^{\alpha\beta} + i \Omega^{ij}\Theta^\alpha_i \dot \Theta^\beta_j\right)\bP_{\alpha\beta} - \frac{1}{2} e\, \left(\bP^2+ m^2\right) \right\} \, , 
\end{equation}
where $\Omega^{ij}$ is the 2nd order antisymmetric invariant tensor of $USp(2n)$;  its inverse $\Omega_{ij}$ will be defined such that
\begin{equation}\label{inverseOm}
\Omega^{ik}\Omega_{ij} = \delta_j^k\, . 
\end{equation} 
The  orthosymplectic phase-space 2-form defined by this action is invertible provided that the mass $m$ is non-zero, and its inverse  gives us the canonical Poisson bracket relations. In particular, one finds this way that\footnote{We shall not need to know  $\{\bX,\bX\}_{PB}$,  which is also non-zero, implying a non-commutative Minkowski spacetime in the quantum theory.}
\begin{equation}\label{ThetaPB}
\left\{\bX^{\alpha\beta},\Theta_i^\gamma\right\}_{PB} =  -\frac{2}{m^2} \bP^{\gamma[\alpha}\Theta_i^{\beta]} \, ,\qquad 
\left\{ \Theta_i^\alpha ,  \Theta_j^\beta \right\}_{PB} =  \frac{2i}{m^2}\Omega_{ij}\bP^{\alpha\beta} \,, 
\end{equation}
where the mass-shell condition has been used to simplify the right hand sides (so one should first replace $m^2$ by $-\bP^2$ before attempting to verify 
Jacobi identities).

The Lorentz Noether charge is now 
\begin{equation}
{\cal J}_{\alpha}{}^{\beta} =  2\bP_{\alpha\gamma}\bX^{\beta\gamma} - \frac{1}{2}\delta_\alpha^\beta\,  \bP_{\gamma\delta}\bX^{\gamma\delta}
- i\Omega^{ij}\Theta^{\gamma}_i \Theta^{\beta}_j \bP_{\alpha\gamma} \, , 
\end{equation}
and the $(n,0)$ supersymmetry charges are 
\begin{equation}\label{Qs}
{\cal Q}^i_\alpha = 2\Omega^{ij} \bP_{\alpha\beta} \Theta^\beta_j \, . 
\end{equation}
As reviewed in the introduction, the massive 6D superparticle with manifest $(n,0)$ supersymmetry actually has $(n,n)$ supersymmetry. 
The $(0,n)$ non-manifest supersymmetry Noether charges are
\begin{equation}\label{nonman}
\tilde{{\cal Q}}_i^{\alpha} = m\, \Theta_i^{\alpha}  \, . 
\end{equation}
Using (\ref{ThetaPB}), one finds that 
\begin{eqnarray}
\left\{ {\cal Q}^i_{\alpha} , \, {\cal Q}^j_{\beta} \right\}_{PB}  &=& -2i\,  \Omega^{ij}\bP_{\alpha\beta}
\,, \nonumber \\
\left\{ \tilde{{\cal Q}}_i^{\alpha} , \, \tilde{{\cal Q}}_j^{\beta} \right\}_{PB} &=& 2i \, \Omega_{ij}\, \bP^{\alpha\beta} \,, \nonumber \\
\left\{ {\cal Q}^i_{\alpha} , \, \tilde{{\cal Q}}_j^{\beta} \right\}_{PB} &=& -im\,  \delta^i_j \delta^{\beta}_{\alpha} \,. 
\end{eqnarray}
One also finds, as expected, that 
\begin{eqnarray}
\left\{{\cal J}_\alpha{}^\beta, {\cal Q}_\gamma^i\right\}_{PB} &=&\  \delta^\beta_\gamma {\cal Q}_\alpha^i - \frac{1}{4} \delta_\alpha^\beta {\cal Q}_\gamma^i \nonumber \\
\left\{{\cal J}_\alpha{}^\beta, \tilde {\cal Q}^\gamma_i\right\}_{PB} &=&  - \delta_\alpha^\gamma \tilde {\cal Q}^\beta_i + \frac{1}{4}\delta_\alpha^\beta \tilde{\cal Q}^\gamma_i\, .
\end{eqnarray}

\subsection{Super-Pauli-Lubanski tensors}

We are now in a position to find supersymmetric analogs of the Pauli-Lubanski  tensors, but we postpone discussion of this issue for $\Xi$ because it is more simply addressed in the supertwistor formulation that we shall be developing later. Written as bi-spinors, the supersymmetric versions of  the PL tensors  (\ref{sigpm}) are 
\begin{equation}\label{sigmas}
\!\!\!\!\!\!\!\!\!\!\!\!  \Sigma^{(+)}_{\alpha\beta} =  {\cal J}_{(\alpha}{}^\gamma{\cal P}_{\beta)\gamma} + \frac{i}{4} \Omega_{ij}{\cal Q}^i_{\alpha}{\cal Q}^j_{\beta}\, , \qquad 
 \Sigma_{(-)}^{\alpha\beta} =  {\cal J}_\gamma^{(\alpha}{\cal P}^{\beta)\gamma} + \frac{i}{4} \Omega^{ij} \tilde{{\cal Q}}_i^{\alpha} \tilde{{\cal Q}}_j^{\beta} \, . 
\end{equation}
One may verify that these bi-spinors have zero Poisson bracket with all supersymmetry charges provided that one makes use of the mass-shell constraint and the relation 
\begin{equation}\label{QtildeQ}
\tilde{\cal Q}_i^\alpha =  -\frac{2}{m}\, \Omega_{ij} \, \bP^{\alpha\beta}{\cal Q}_\beta^j \, , 
\end{equation}
which is valid for the superparticle as a consequence of the expressions (\ref{Qs}) and (\ref{nonman}) for the supercharges in terms of the phase superspace coordinates. 

A clarification is in order here. The existence of the ``hidden'' $(0,n)$ supersymmetry charges is a special feature of the superparticle model under study. Should it not be possible to define
super-PL tensors for $(n,0)$ supersymmetry that involve only the $(n,0)$ supercharges? The answer is a qualified yes. If our  interest is in the quadratic Casimir of the $(n,0)$ supersymmetry 
algebra that generalizes the usual $\Sigma^2$ invariant of the Poincar\'e algebra (for example) then we may proceed by defining the new traceless bi-spinor
\begin{equation}\label{upsilon}
 \Upsilon_\alpha{}^\beta = {\cal P}^{\beta\gamma}\Sigma^{(+)}_{\alpha\gamma} = \frac{1}{2} {\cal P}^{\beta\gamma}\left({\cal P}_{\alpha\delta} {\cal J}_\gamma{}^\delta + \frac{i}{2}\Omega_{ij}{\cal Q}_\alpha^i {\cal Q}_\gamma^j\right) - \frac{1}{8} {\cal P}^2 {\cal J}_\alpha{}^\beta\, . 
 \end{equation}
This is equivalent to a 2nd-rank antisymmetric tensor, or 2-form, and hence also to a 4-form (the relevance of this observation will be apparent shortly).  It has zero Poisson bracket
with the ${\cal Q}$ supercharges, so its norm $\Upsilon^2 \equiv \Upsilon_\alpha{}^\beta\Upsilon_\beta{}^\alpha$ is a super-Poincar\'e invariant.   This constructs a Casimir from 
$\Sigma^{(+)}$ and ${\cal P}$,  valid for the $(n,0)$ supersymmetry algebra; this  is possible because in 6D we can decompose $\Sigma$ into its self-dual and anti-self-dual parts. 
In other spacetime dimensions there is no analog of the strictly $(n,0)$ super-extension of $\Sigma$  that commutes with the $(n,0)$ supercharges (and the same is true for $\Xi$ even in 6D). 
The standard resolution of this problem, for the $(D-3)$ form $\Sigma$, relies on the fact that there {\it is} a super-invariant extension of the  $(D-2)$-form  found by taking  the exterior product of  a 
$\Sigma$ with ${\cal P}$; see e.g.  \cite{Buchbinder:1998qv} for the 4D case, which was generalized  to arbitrary spacetime dimension in \cite{Zumino:2004nb}.  In 6D this 4-form is 
precisely our  $\Upsilon$. 

We have still to address the issue of the relation between $\Upsilon^2$ and $\Sigma^2$. Recall that $\Sigma^2$ is a contraction of  $\Sigma^{(+)}$ with $\Sigma_{(-)}$, but the definition of the latter 
in (\ref{sigmas})  involves the hidden $(0,n)$ supersymmetries; moreover, one needs the superparticle mass-shell condition and the relation (\ref{QtildeQ}) between the $(n,0)$ and $(0,n)$ supercharges to show that $\Sigma_{(-)}$ has zero Poisson bracket with  the $(n,0)$ supercharges. This makes it appear that $\Sigma^2$ is defined only for the superparticle. However, if we use the relation (\ref{QtildeQ}) to rewrite  $\Sigma_{(-)}$ in terms of the $(n,0)$ supercharges, then we find that 
\begin{equation}
{\cal P}_{\alpha\gamma} \Sigma_{(-)}^{\beta\gamma} =   {\cal P}^{\beta\gamma}\Sigma^{(+)}_{\alpha\gamma}  \equiv \Upsilon_\alpha{}^\beta \, . 
\end{equation}
This shows that the first of the identies of (\ref{PLids}) remains valid for the super-PL tensors as we have defined them.  A corollary is that 
\begin{equation}
\Upsilon^2 = - \frac{1}{4}{\cal P}^2 \Sigma^2 \, . 
\end{equation}
What this shows is that the scalar $\Sigma^2$,  constructed as a Casimir for the  $(n,n)$ supersymmetry algebra of the superparticle  is valid in full generality when considered as a Casimir for massive representations (${\cal P}^2=-m^2\ne0$) of just the $(n,0)$  supersymmetry algebra.

\section{Twistor formulation of massive 6D particle}
\setcounter{equation}{0}

We can solve the mass-shell constraint $\bP^2+m^2=0$ of the action (\ref{phase})  by first  setting
\begin{equation}\label{Psq1}
\bP_{\alpha\beta} = \frac{1}{2} \bU_\alpha^I \bU_\beta^J \, \Omega_{JI} \, , 
\end{equation}
where $\bU$ is a $USp(4)$ 4-plet $(I=1,2,3,4)$ of  $SU^*(4)$ spinors, and  $\Omega$ is here the standard invertible antisymmetric invariant tensor of $USp(4)$. Then, 
viewing $\bU$ as the $4\times 4$ matrix with entries $\bU_\alpha^I$, we impose the constraint
\begin{equation}\label{massS}
0= \varphi := \det \bU +m^2 \, . 
\end{equation} 
To verify that this solves the mass-shell constraint one needs the identity
\begin{equation}
3\Omega_{I[J} \Omega_{KL]} =  \epsilon_{IJKL}\, , 
\end{equation}
where $\epsilon_{IJKL}$ is the $USp(4)$ invariant alternating tensor. 
A corollary of (\ref{Psq1}) is that 
\begin{equation}
\left(\bP^2\right)^2 = 16\det \bP = (\det \bU)^2 \quad \Rightarrow \quad \det \bU = \pm \, \bP^2\, .
\end{equation}
where the first equality is from (\ref{detp}). 
Choosing the upper sign for compatibility with (\ref{massS}), we see that the  constraint $\varphi=0$  is just the original mass-shell constraint in spinor form!  
Notice that the solution (\ref{Psq1}) of the original mass-shell constraint is invariant under {\it local}  $USp(4)$ transformations, so we can anticipate that new constraints associated to a new $USp(4)\cong {\rm Spin}(5)$ gauge invariance will emerge. 

Substitution for $\bP$ gives
\begin{equation}\label{defW6}
\dot{\bX} \cdot \bP  = {\bU}_\alpha^I \dot{\bW}^\alpha_I + \frac{d}{dt} \left(\dots\right) \, , \qquad \bW^\alpha_I= X^{\alpha\beta}\bU_\beta^J \Omega_{JI}\, . 
\end{equation}
Let us define 
\begin{equation}
\Lambda^{IJ}= \bU_\alpha^{(I} \bW^{\alpha J)}\, , \qquad \bW^{\alpha J}= \Omega^{JK} \bW^\alpha_K \, , 
\end{equation}
where $\Omega^{IJ}$ is defined (as for $\Omega^{ij}$) such that 
\begin{equation}
\Omega^{IK}\Omega_{JK} = \delta^I_J\,  .  
\end{equation}
In general, we use $\Omega_{IJ}$ ($\Omega^{IJ}$) to lower (raise) $USp(4)$ indices according to the convention (for arbitrary $USp(4)$ 4-plet $Z$)  that
\begin{equation} 
Z^I = \Omega^{IJ}Z_J\, , \qquad Z_I = Z^J \Omega_{JI}\, , 
\end{equation}
from which it follows that
\begin{equation}
\Omega_I{}^J = \delta_I^J = - \Omega^I{}_J \, . 
\end{equation}

Given the definition of $\bW^\alpha_I$, we have $\Lambda^{IJ}\equiv0$, so this must be imposed as a set of constraints when $\bW^\alpha_I$ is considered as a set of  independent variables. These 
are the ``spin-shell'' constraints; this terminology will be justified shortly. Imposing these  constraints by Lagrange multipliers $s_{IJ}$ and the new mass-shell constraint by a Lagrange multiplier $\rho$, 
we arrive at the following twistor form of the action for a massive 6D particle:
\begin{equation}\label{6Dact}
S= \int\! dt \left\{ \bU_\alpha^I \dot{\bW}^\alpha_I  - s_{IJ} \Lambda^{IJ}  - \rho\, \varphi \right\}\, . 
\end{equation}
The constraint functions  $\Lambda^{IJ}$ generate the expected local $USp(4)$ gauge transformations, via the canonical Poisson bracket relations
\begin{equation}\label{PB6D}
\left\{\bW^\beta_J ,  \bU_\alpha^I \right\}_{PB} = \delta_\alpha^\beta \delta^I_J \, . 
\end{equation}
Since $\det \bU$ is manifestly $USp(4)$ gauge invariant, the additional constraint function has zero Poisson bracket with $\Lambda^{IJ}$, and hence all constraints are first class. 

As a consistency check, let us verify that the physical phase space dimension is unchanged by the process that converts the standard massive particle action into the new twistor action.  We started with a phase space of dimension $2\times 6=12$ subject to a single first-class constraint, implying a physical phase space of dimension $12-2=10$. We now have a phase space of (real) dimension $2\times (4\times4)=32$ subject to $10+1=11$ first-class constraints, implying a physical phase space dimension of $32-22=10$. 

\subsection{Gauge invariances}

The constraint functions $\Lambda^{IJ}$ generate the ${\rm Spin}(5)$ gauge transformations of the canonical variables, which are
\begin{equation}
\delta_\ell \bU_\alpha^I = -\bU_\alpha^J \ell_J{}^I \, , \qquad \delta_\ell \bW^\alpha_I = \ell_I{}^J \bW^\alpha_J  \qquad (\ell^{IJ} = \ell^{JI})\, . 
\end{equation}
This is an invariance of the action provided that we assign the following gauge transformation to the Lagrange multiplier
\begin{equation}
\delta_\ell s_I{}^J= \dot\ell_I{}^J + \ell_I{}^K s_K{}^J - s_I{}^K \ell_K{}^J\, . 
\end{equation}
This ${\rm Spin}(5)$ gauge invariance is expected because it was introduced when we solved the mass-shell constraint $\bP^2+m^2=0$, but 
what is the significance of the additional gauge invariance associated to the constraint $\varphi=0$?  

To answer this question, we begin by observing that the additional  non-zero gauge transformations are 
\begin{equation}
\delta_\lambda \bW^\alpha_I = \lambda m \bV^\alpha_I\, , \qquad \delta \rho= \dot \lambda
\end{equation}
where $\lambda(t)$ is the infinitesimal parameter, and 
\begin{equation}
\bV_I^{\alpha} = \frac{1}{6m} \epsilon_{IJKL}\epsilon^{\alpha\beta\gamma\delta}\bU^J_{\beta}\bU^K_{\gamma}\bU^L_{\delta} \, . 
\end{equation}
This new opposite-chirality commuting spinor variable is essentially the inverse of $\bU$  on the surface $\varphi=0$ since, on this surface, 
\begin{equation}
\bV^\alpha_I \bU_\alpha^J  = -m\delta_I^J\, , \qquad \bV^\alpha_I \bU_\beta^I = -m\delta^\alpha_\beta\, \qquad (\det \bU =\det\bV = -m^2)\, . 
\end{equation}
A useful identity  is 
\begin{equation}
\varepsilon^{\alpha\beta\gamma\delta} \bU_\gamma^I \bU_\delta^J \equiv - \varepsilon^{IJKL} \bV_K^\alpha\bV_L^\beta \qquad (\varphi=0)\, . 
\end{equation}
This allows us to express $\bP$ on the $\varphi=0$  surface  as 
\begin{equation}
\bP^{\alpha\beta} = -\frac12\, \bV_I^{\alpha}\bV_J^{\beta} \Omega^{JI}  \qquad (\varphi=0)\, . 
\end{equation}

Next, we observe that we may add to any gauge transformation the following ``trivial'' gauge transformation with parameter $\xi(t)$:
\begin{eqnarray}
\delta_\xi \bU_\alpha^I &=& - \xi \frac{\delta S}{\delta \bW^\alpha_I} = \xi \left(\dot{\bU}_\alpha^I - \bU_\alpha^J s_J{}^I\right)\, , \nonumber \\
\delta_\xi\bW^\alpha_I &=& \xi \frac{\delta S}{\delta \bU_\alpha^I} = \xi\left(\dot{\bW}^\alpha_I + s^I{}_J\bW^\alpha_J -m\rho \bV^\alpha_I\right)\, . 
\end{eqnarray}
This is manifestly a gauge invariance, but a ``trivial'' one  because the transformations are zero on solutions of the equations of motion. Now consider the linear combination 
\begin{equation}
\delta'_\xi = \delta_\xi + \delta_\lambda+ \delta_\ell\, , \qquad \lambda= \rho\,\xi\, , \quad \ell_I{}^J = s_I{}^J \xi\, . 
\end{equation}
One finds that the $\delta'_\xi$ transformations of the canonical variables are those due to a reparametrization of the worldline time:
\begin{equation}
\delta'_\xi \bU_\alpha^I = \xi\, \dot{\bU}_\alpha^I\, , \qquad \delta'_\xi \bW^\alpha_I = \xi\, \dot{\bW}^\alpha_I\, . 
\end{equation}
We conclude that the additional constraint is associated with the time reparametrization invariance of the action, as expected from its equivalence to 
the original mass-shell constraint. 

\subsection{Poincar\'e invariance}

In the new spinor variables, the Poincar\'e Noether charges are
\begin{equation}
{\cal P}_{\alpha\beta} = \frac12 \bU^I_{\alpha}\bU^J_{\beta}\Omega_{JI} \,, \qquad 
{\cal J}_{\alpha}{}^{\beta} = \bU^I_{\alpha}\bW^{\beta}_I - \frac{1}{4}\delta_{\alpha}^{\beta} \, (UW)\, , 
\end{equation}
where we use the shorthand notation 
\begin{equation}
(UW) \equiv  \bU^I_{\gamma}\bW_I^{\gamma} \,.
\end{equation}
Using these expressions in (\ref{sigmas}), and the constraint $\det \bU=-m^2$, we find that 
\begin{eqnarray}\label{sigs}
\Sigma^{(+)}_{\alpha\beta} = \frac12 \bU^I_{\alpha}\bU^J_{\beta}\Lambda_{IJ} \,, \qquad 
\Sigma_{(-)}^{\alpha\beta}  = \frac12  \bV_I^{\alpha} \bV_J^{\beta} \Lambda^{IJ} \,,
\end{eqnarray}
From (\ref{XS}) it then follows that 
\begin{equation}\label{spinor-Xi}
\Xi_{\alpha\beta}= \bU_\alpha^J \bU_\beta^K \Lambda_K{}^I \Lambda_{IJ} - \frac{1}{2} \bP_{\alpha\beta} \Lambda_K{}^L\Lambda_L{}^K\, . 
\end{equation}
Notice that 
\begin{equation}
\Sigma^2 = -\frac{m^2}{4} \Lambda_I{}^J \Lambda_J{}^I\, . 
\end{equation}
The left hand side is  a Poincar\'e Casimir while the right hand side is proportional to the quadratic Casimir of the spin-shell group. 
This generalizes to 6D the observation for the massive 4D particle in \cite{Mezincescu:2013nta}, but the connection between the spin-shell constraints and the particle's 
spin is already evident from the expressions (\ref{sigs}) and (\ref{spinor-Xi})  because they show that all PL tensors are zero on spin-shell,  and this tells us that the particle 
has zero spin.

\section{Supertwistors and the massive 6D  superparticle}
\setcounter{equation}{0}

We now turn to the massive superparticle with action (\ref{saction}), which has manifest $(n,0)$ supersymmetry, and we solve the mass-shell 
constraint as in (\ref{Psq1}). As before this leads to the new mass-shell constraint $0=\det U +m^2 \equiv \varphi$.  Substitution for $\bP$ as before 
now leads to
\begin{equation}
\left( \dot{\bX}^{\alpha\beta} + i \Omega^{ij}\Theta_i^{\alpha}\dot \Theta_j^{\beta} \right) P_{\alpha\beta} = 
\bU_{\alpha}^I \dot{\bW}^{\alpha}_I + \frac{i}{2}\Omega^{ij}\Omega_{JI} \, \mu_i{}^I \dot \mu_j{}^J  + \frac{d}{dt}\left(\ldots\right) \,,
\end{equation}
where 
\begin{equation}
\bW^{\alpha}_I = \left(\bX^{\alpha\beta}\bU^J_{\beta} - \frac{i}{2} \Omega^{ij}\Theta^{\alpha}_i \mu_j{}^J\right) \Omega_{JI} \,, \qquad 
\mu_i{}^I = \bU^I_{\alpha} \Theta_i^{\alpha} \,.
\end{equation}
The definition of $\bW$ leads to the identity
\begin{equation}\label{ssshell}
0 \equiv  \bU_\alpha^{(I} \bW^{\alpha J)} - \frac{i}{2} \Omega^{ij} \mu_i{}^I \mu_j{}^J \equiv \Lambda^{IJ}\, . 
\end{equation}
As before, to promote  $\bW$ to an independent variable we must impose this identity as a constraint, so the action in the new variables is 
\begin{equation}\label{6Dsact}
S= \int\! dt \left\{ \bU_\alpha^I \dot{\bW}^\alpha_I  + \frac{i}{2}\mu^i{}_I \,  \dot \mu_i{}^I - s_{IJ} \Lambda^{IJ}  - \rho\, \varphi \right\}\, ,  
\end{equation}
where 
\begin{equation}
\mu^i{}_I  =\Omega^{ij}\, \mu_j{}^J\Omega_{JI}\, . 
\end{equation}
The phase-space variables are the components of a pair  of  6D supertwistors ($I=1,2,3,4$ rather than $I=1,2$) but the 6D superconformal invariance 
is broken by the $\varphi=0$ constraint. 

The new superparticle action (\ref{6Dsact}) is manifestly Lorentz invariant, with Noether charges
\begin{equation}
{\cal J}_{\alpha}{}^{\beta} = \bU^I_{\alpha}\bW^{\beta}_I - \frac{1}{4}\delta^{\beta}_{\alpha}\left(UW\right) \,. 
\end{equation}
There is no fermion bilinear term, as could have been anticipated from the fact  that the anticommuting variables $\mu_i{}^I$ are now Lorentz scalars. 
The action is also invariant under all $(n,n)$ supersymmetries, with Noether charges are
\begin{equation}
{\cal Q}^i_{\alpha} = \bU^I_{\alpha}\mu^i{}_I \, , \qquad \tilde{\cal Q}_i^\alpha = - \mu_i{}^I \bV^\alpha_I \, . 
\end{equation}
This may be verified using  the Poisson bracket relation (\ref{PB6D}) and the new (symmetric) Poisson bracket relations
\begin{equation}
\left\{ \mu^i{}_I, \mu_j{}^J\right\}_{PB} \equiv \left\{\mu_j{}^J,\mu^i{}_I\right\}_{PB} =-i \delta^i_j \delta_I^J\, . 
\end{equation}
In particular, the spin-shell constraints are $(n,n)$ supersymmetric because
\begin{equation}
\left\{{\cal Q}_\alpha^k,\Lambda^{IJ}\right\}_{PB} = 0\, , \qquad \left\{\tilde{\cal Q}^\alpha_k,\Lambda^{IJ}\right\}_{PB} = 0\, . 
\end{equation}

Using the supertwistor expressions for the  super-Poincar\'e charges in the expressions (\ref{sigmas}) for the super-Pauli Lubanski 3-form $\Sigma$, we find that 
\begin{eqnarray}\label{super-sigs}
\Sigma^{(+)}_{\alpha\beta} = \frac12 \bU^I_{\alpha}\bU^J_{\beta}\Lambda_{IJ} \,, \qquad 
\Sigma_{(-)}^{\alpha\beta}  = \frac12  \bV_I^{\alpha} \bV_J^{\beta} \Lambda^{IJ} \,. 
\end{eqnarray}
Formally, this is {\it identical} to the result that we found for the bosonic particle; the only difference is that the spin-shell constraint functions, 
given by (\ref{ssshell}), now include terms bilinear  in the anticommuting variables $\mu_i{}^I$.  This result should not be a surprise because 
the spinor variables $\bU$ are inert under supersymmetry and, as we have just seen, the superparticle extension of the spin-shell constraint functons 
are supersymmetric. 

It is now obvious how to find the supersymmetric extension of the Pauli-Lubanski  vector $\Xi$ of (\ref{Xi}). We just return to the twistor expression (\ref{spinor-Xi})
and re-interpret $\Lambda^{IJ}$  as the superparticle  spin-shell constraint functions. This gives us 
\begin{equation}\label{spinor-Xi2}
\Xi_{\alpha\beta}= \bU_\alpha^J \bU_\beta^K \Lambda_K{}^I \Lambda_{IJ} - \frac{1}{2} \bP_{\alpha\beta}\,  \Lambda_K{}^L\Lambda_L{}^K\, ,  
\end{equation}
where $\Lambda^{IJ}$ are now the superparticle spin-shell contraint functions.  

\subsection{Quantum theory}

If we define a massive particle of zero superspin to be one for which all super-PL tensors are zero, then the spin-shell constraints of the massive superparticle
tell us that it has zero superspin.  The canonical  anticommutation relations of the $8n$ fermionic phase-space variables of the action (\ref{6Dact}) are
\begin{equation}
\left\{ \mu^i{}_I, \mu_j{}^J\right\} = \delta^i_j \delta_I^J\, . 
\end{equation}
This implies a supermultiplet with $2^{4n}$ independent polarization states. For $n=1$  this gives us a massive supermultiplet with $16$ components,  
and zero superspin tells us that this must be the 6D Proca multiplet, for which the bosonic content is one massive vector and three scalar fields. This is 
a massive supermultiplet of $(1,0)$ 6D supersymmetry. If we declare the particles of this supermultiplet to carry a central charge, which can be done by allowing 
superparticle wavefunction to be complex, then it is also  a supermultiplet of $(1,1)$ 6D supersymmetry, with a central charge saturating the 
BPS unitarity bound implied by supersymmetry. 

In other words,  we have the choice of quantizing preserving only the manifest $(1,0)$ 6D supersymmetry, in which case 
we can impose a reality condition on the superparticle wavefunction, so as to get the Proca  supermultiplet, or we can insist on  preserving the full $(1,1)$ 6D supersymmetry,
in which case we get a pair of Proca supermultiplets with equal and opposite central charges. The latter supermultiplet is exactly what one gets by keeping a single massive level of the 
Kaluza-Klein tower resulting from toroidal compactification to 6D of the 10D Maxwell supermultiplet.

\section{Discussion}
\setcounter{equation}{0}

In the twistor formulation of particle mechanics, in $D$ spacetime dimensions, the usual mass-shell constraint is solved by expressing the $D$-momentum
as a bi-spinor. The spinor variable introduced by this solution is then viewed as a new phase-space coordinate, and its canonical conjugate is another spinor. 
Taken together these canonically conjugate spinors constitute a twistor, a spinor of the conformal group.  However, for this construction to work, it must be
that the physical phase space has the same dimension as it did originally, and this is a significant constraint. 

For $D=3,4,6$ we have $D=2+ K$, where $K$ is the dimension  (over $\bR$) of $\bK=\bR,\bC,\bH$ (the reals, complex numbers and quaternions), and a minimal spinor is a doublet 
of  $Sl(2;\bK)$; in addition, a set of $N$ such spinors is an $N$-plet of the internal symmetry group $U(N;\bK)$ \cite{Kugo:1982bn}. Since a twistor comprises a pair of spinors, each of which has
$2N$ $\bK$-valued components, the total dimension over $\bR$ of the vector space spanned by $N$ twistors is $4NK$. However, 
since\footnote{
The dimension is over $\bR$, and we use  the fact that  $U(n;\bK)$ is isomorphic to  $O(n)$, $U(n)$, $USp(2n)$ for $\bK=\bR,\bC,\bH$, respectively.}
\begin{equation}
2\, {\rm dim}\,  U(N;\bK) = N(N+1)K -2N\, , 
\end{equation}
the combined effect of  $U(N;\bK)$ spin-shell constraints and the associated $U(N;\bK)$  gauge-invariance is to reduce the  phase space to one with  dimension  
$2N-N(N-3)K$. On the other hand, the physical phase dimenson is
$2(D-1)=2(1+K)$. This means that $2(N-1)= (N-1)(N-2)K$, assuming the absence of any constraints other than the spin-shell constraints; allowing for the possibility of additional constraints 
we thus arrive at the inequality
\begin{equation}\label{inequality}
(N-1)\left[2-(N-2)K\right] \ge0\, . 
\end{equation}

For the twistor form of the massless point particle in dimensions $D=3,4,6$ we need $N=1$, in which case the above inequality is saturated. 
The massive particle requires both $N>1$ and at least one additional constraint (in order to solve the mass-shell condition) and this is compatible with the above inequality only for $N=2$,  in which case (\ref{inequality}) is satisfied with  
the left hand side of (\ref{inequality}) equal to $2$.  This allows either two additional second-class constraints or one additional first-class constraint but, as we explain below, 
the twistor form of the massive particle must have one additional first-class constraint. These conditions are indeed  realized by  the double-twistor formulation of the massive particle, as we have shown here for $D=6$. Our result thus complements and completes  earlier work on twistor constructions of this general type.  

One may ask why there is an additional constraint for the massive particle. Actually, one should expect an additional constraint because of the worldline time
reparametrization invariance  of the action, so what has to be explained is why no such additional constraint is needed for the massless particle. The answer is that
in the massless case, but not in the massive case, one can combine a time-reparametrization with a spin-shell gauge transformation to arrive at a ``trivial'' gauge 
transformation: one for which the transformations are all zero for solutions of the equations of motion. As such gauge transformations have no physical effect, 
time-reparametrization invariance is not independent of the spin-shell gauge invariance for a massless particle. For a massive particle the equations of motion differ, such that 
the spin-shell  constraint functions  no longer suffice to generate all non-trivial gauge transformations, so an additional constraint associated to time reparametrization 
invariance is required.  

Another way to see how the possibilities for a twistor formulation of particle mechanics are limited is no notice that there must be a coincidence (or near coincidence) 
between the spin-shell group  $U(N;\bK)$  and  Wigner's  ``little group''  (the subgroup of the Poincar\'e group relevant to the classification  of elementary particles) with 
$N=1$ applying to massless particles  and $N=2$ to massive particles. The reason is that the Pauli-Lubanski spin tensors, which are identically zero when expressed in terms of the usual 
phase space variables of a spinless particle, are zero when expressed in twistor variables only as a consequence of the spin-shell constraints. Consequently, 
the little-group generators become identified with the spin-shell group generators in  a standard Lorentz frame. The  massive 4D particle is a mild exception to this rule because 
the spin-shell group is $U(2)$ but the rotation group is $SU(2)$ (a ``near coincidence''); however,  the $U(1)$ factor drops out of the Pauli-Lubanski vector, which becomes identified with the generators of  space rotations. For the massive 6D particle considered here, the spin-shell group is $USp(4)\cong {\rm Spin}(5)$, which has the same Lie algebra as the rotation group, 
and the Pauli-Lubanski 3-form is equivalent in a standard Lorentz frame  to the adjoint ${\bf 10}$ of the ${\rm Spin}(5)$ algebra, spanned by the spin-shell constraint functions. 

In addition to finding the twistor formulation of the massive 6D particle, we have extended the construction to a supertwistor formulation of the massive superparticle. A nice feature
of this construction  is that it makes manifest the full supersymmetry invariance, which is always that of a BPS superparticle with $(n,n)$ supersymmetry for some $n$ 
Exactly the same action would result from a supertwistor reformulation of the  ``kappa-symmetric'' BPS superparticle action for which the $(n,n)$ supersymmetry is manifest from 
the start. This  follows from the general arguments of \cite{Mezincescu:2014zba}, summarised in the introduction, but it was also verified explicitly for $D=4$ in \cite{Mezincescu:2013nta}.

Implicit in our results is a supertwistor formulation of the {\it massless} 6D superparticle with $(n,n)$  supersymmetry, obtained by setting $m=0$. Notice that this massless superparticle action
cannot be equivalent to the standard massless superparticle action with manifest $(n,0)$ supersymmetry because (and in contrast to the massive case) the latter does not have a hidden 
$(0,n)$ supersymmetry. Also, there is no 
previously known supertwistor formulation of the massless $(n,n)$-supersymmetric superparticle (only the $(n,0)$ cases are known). We suspect that our indirect solution to this problem 
is not the most economical one, but we have not investigated this. 

The spin-content of any relativistic particle mechanics model is determined by the Pauli-Lubanski  (PL) tensors (which are functions on phase space in the context of classical particle mechanics).
All PL tensors are zero for a  massive particle of zero spin; for the twistor form of the particle's action this is true as a consequence of the spin-shell  constraints (hence the terminology). 
We have established a similar result here for the supertwistor form of the massive 6D superparticle: all super-PL tensors are zero as a consequence of the spin-shell constraints. 
In the quantum theory this implies that the superparticle describes a 6D supermultiplet of zero superspin. In the simplest ($n=1$) case this is the 6D Proca supermultiplet for a massive vector 
field, three scalar fields and their spin-$1/2$ superpartners, which must be centrally charged if we insist on quantizing preserving the full $(1,1)$ supersymmetry. 

Our construction of the super-PL tensors differs from the standard one. In fact, this terminology is not  used in the standard construction of super-Poincar\'e Casimirs, for good reason. For example, for  $D=4$ there is no ${\cal N}=1$ supersymmetric extension of the usual Pauli-Lubanski spin-vector that commutes with the supersymmetry generator.  The problem is milder in 6D because
of special features of this dimension (one can use only the self-dual part of $\Sigma$) but it is still true that not all 6D PL tensors  have a strictly $(n,0)$ extension that commutes with the $(n,0)$ supercharges.  In 4D  this problem is solved  by the existence of a supersymmetric extension of the  antisymmetric tensor constructed by taking the exterior product of the momentum generator with the  PL spin-vector. The same construction 
can be used in 6D, and higher dimensions, but  the method  has not yet been developed  so that it applies to {\it all} super-Poincar\'e Casimirs.  Our  superparticle approach provides an alternative route to the construction of super-Poincar\'e Casimirs: by taking account of the ``hidden'' supersymmetries of the superparticle model \cite{Mezincescu:2014zba}, we find a super-PL tensor invariant under all supersymmetries. We have shown for the simplest case how the scalars constructed from these super-PL tensors become model-independent Casimirs for the manifest supersymmetry algebra.  We suspect that this idea could lead to a simple general construction  of all super-Poincar\'e Casimirs, but we leave this to the future.

\subsection{\bR\,\bC\,\bH\,\bO}

The original suggestion of a close relationship between (Minkowski space) supersymmetry in spacetime dimensions $D=3,4,6,10$ and the division algebras  was based partly on the coincidence that the double cover of the Lorentz group in these dimensions is $Sl(2;\bK)$ for $\bK=\bR,\bC,\bH,\bO$ \cite{Kugo:1982bn},  as confirmed by Sudbery  for $\bK=\bO$ by a suitable definition of $Sl(2;\bO)$  \cite{Sudbery:1983}.  The results reported here provide further evidence of  this relationship for the $D=6$ case, as would be manifest if we had used  $2$-component quaternionic spinors instead of $4$-component complex spinors;  indeed, a quaternionic formulation of the massless $D=6$ superparticle (although not its supertwistor version)  was worked out  in  \cite{Kimura:1988ta}.  

The work reported here is potentially of relevance to the massless $D=10$ superparticle because the massive 6D superparticle can be viewed as a massless 10D particle in a spacetime that is a product of 6D Minkowski space with a 4-torus,  with a fixed non-zero $4$-momentum on the 4-torus. This is easily seen from the usual phase-space formulation of the massive 6D superparticle but it is not at all obvious from its supertwistor phase-space formulation.  If  this 10D origin could be understood in 6D twistor terms, it could provide a clue to some novel reformulation of the 10D massless superparticle.  

We should point that there is already an octonionic formulation of the massless 10D superparticle \cite{Oda:1988sz,Schray:1994fc}, and a twistor version of it was proposed in \cite{Cederwall:1992bi}. 
Another  $D=10$ result involving both the octonions and twistors was presented in \cite{Galperin:1992pz}: the super-Maxwell field equations for $D=3,4,6,10$ can be solved (by a twistor transform) in  terms of a $\bK$-valued  worldline superfield satisfying a ``$\bK$-chiral'' constraint.

Another obvious question is whether the results reported here for the massive 6D superparticle could be generalised to 10D; i.e. to the {\it massive}  $D=10$ superparticle with $(1,0)$ supersymmetry. This would be of great interest  because the action actually has $(1,1)$ 10D supersymmetry and is just a gauge-fixed version of the D0-brane action of IIA superstring theory \cite{Mezincescu:2014zba}. However,  we have nothing definite to say about this case, and so leave it to future investigations. 

\section*{Acknowledgements} We thank Luca Mezincescu for bringing ref. \cite{Zumino:2004nb} to our attention, and Martin Cederwall for helpful correspondence. PKT acknowledges support from the UK Science and Technology Facilities Council (grant ST/L000385/1). AJR is supported by a grant from the London Mathematical Society, and he thanks the University of Groningen for hospitality during the writing of this paper.

\bigskip


\providecommand{\href}[2]{#2}\begingroup\raggedright\endgroup

\end{document}